# Crossover of Ising- to Rashba-Type Superconductivity in Epitaxial Bi$_2$Se$_3$/Monolayer NbSe$_2$ Heterostructures

Hemian Yi[1,7], Lun-Hui Hu[1,7], Yuanxi Wang[1,2,7], Run Xiao[1], Jiaqi Cai[3], Danielle Reifsnyder Hickey[4,5], Chengye Dong[5], Yi-Fan Zhao[1], Ling-Jie Zhou[1], Ruoxi Zhang[1], Anthony R. Richardella[1], Nasim Alem[5], Joshua A. Robinson[5], Moses H. W. Chan[1], Xiaodong Xu[3,6], Nitin Samarth[1], Chao-Xing Liu[1], and Cui-Zu Chang[1]

[1]Department of Physics, The Pennsylvania State University, University Park, PA 16802, USA

[2]Department of Physics, University of North Texas, Denton, TX 76203, USA

[3]Department of Physics, University of Washington, Seattle, WA 98195, USA

[4]Department of Chemistry, The Pennsylvania State University, University Park, PA 16802, USA

[5]Department of Materials Science and Engineering, The Pennsylvania State University, University Park, PA 16802, USA

[6]Department of Materials Science and Engineering, University of Washington, Seattle, WA 98195, USA

[7]These authors contributed equally: Hemian Yi, Lun-Hui Hu, Yuanxi Wang

Corresponding author: cxc955@psu.edu (C.-Z. C.)

**Abstract: A topological insulator (TI) interfaced with an *s*-wave superconductor has been predicted to host an unusual form of superconductivity known as topological superconductivity (TSC)[1]. Molecular beam epitaxy (MBE) has been the primary approach in the scalable synthesis of the TI/superconductor heterostructures. Although the growth of epitaxial TI films on *s*-wave superconductors has been achieved[2-4], it remains an outstanding challenge for synthesizing atomically thin TI/superconductor heterostructures, which are** **critical for engineering the TSC phase. Here, we used MBE to grow $Bi_2Se_3$ films with the controlled thickness on monolayer $NbSe_2$ and performed *in-situ* angle-resolved photoemission spectroscopy and *ex-situ* magneto-transport measurements on these $Bi_2Se_3$/monolayer $NbSe_2$ heterostructures. We found that the emergence of Rashba-type bulk quantum well bands and spin-nondegenerate surface states coincides with a marked suppression of the in-plane upper critical magnetic field of the superconductivity in $Bi_2Se_3$/monolayer $NbSe_2$ heterostructures. This is the signature of a crossover from Ising- to Rashba-type superconducting pairings, induced by altering $Bi_2Se_3$ film thickness. Our work opens a new route for exploring a robust TSC phase in TI/Ising superconductor heterostructures.**

**Main text:** Since the discovery of heavy-fermion superconductors and high-temperature superconductors in cuprate-perovskite ceramics, research on unconventional (i.e. non-*s*-wave) superconductivity continues to reveal new questions and challenges. An exciting possibility is topological superconductivity (TSC) in hybrid structures where a topological insulator (TI) is proximally coupled to a conventional *s*-wave superconductor[1]. A primary barrier to the growth of TI films on *s*-wave superconductors (or vice versa) lies in the difficulty of achieving an atomically sharp interface due to a host of experimental challenges such as chemical reactivity and uncontrolled nucleation. To date, the most successful realization of TIs interfaced with *s*-wave superconductors is the molecular beam epitaxy (MBE)-grown $Bi_2Se_3$ or $Bi_2Te_3$ films on bulk crystals of superconducting $NbSe_2$ (Refs.[2-6]). However, for device fabrication and potential application in topological quantum computations, MBE-grown $NbSe_2$ films are preferred over $NbSe_2$ bulk crystals.

Prior studies have demonstrated that the superconductivity in atomically thin, particularly monolayer NbSe$_2$ films persists under in-plane magnetic fields far beyond the Pauli limit ($H_p$) due to its Ising-type superconducting pairing symmetry[7,8]. In Ising-type superconductors, electron spins are pinned to the out-of-plane direction due to the breaking of SU(2) spin rotation via the strong Ising-type spin-orbit interaction. This property makes the breaking of Cooper pairs difficult under an in-plane magnetic field. The TSC phase has also been theoretically predicted in Ising-type superconductors, particularly in transition metal dichalcogenides [9-11]. We may anticipate that a robust TSC phase appears in a TI/Ising superconductor (e.g. Bi$_2$Se$_3$/monolayer NbSe$_2$) heterostructure due to its complex spin-orbit coupling form. However, experimental studies along this direction are to date lacking, primarily because the Ising-type superconductor becomes non-superconducting after being covered by a TI film.

In this work, we used MBE to fabricate Bi$_2$Se$_3$/monolayer NbSe$_2$ heterostructures with different Bi$_2$Se$_3$ thicknesses and performed *in-situ* angle-resolved photoemission spectroscopy (ARPES) and *ex-situ* electrical transport on these heterostructures. We observed that the Dirac surface states (SSs) emerge when the thickness of Bi$_2$Se$_3$ is greater than 3 quintuple layers (QLs), which is reduced compared with previous studies of thicker (i.e. ≥6 QL) Bi$_2$Se$_3$ films grown on sapphire[12], bilayer graphene[13], and Si(111) (Ref.[14]) substrates. We found that Rashba-type conduction bands and spin-nondegenerate Dirac SSs appear when the Bi$_2$Se$_3$ thickness is greater than 2 QL. These features observed in ARPES studies were found to agree well with our first-principles calculations. Moreover, we found that the in-plane upper critical magnetic field (i.e., $H_{c2,\parallel}/H_P$) of the superconducting Bi$_2$Se$_3$/monolayer NbSe$_2$ heterostructures is greatly reduced when the Rashba-type bulk conduction bands and Dirac SSs emerge, implying the crossover from the Ising- to Rashba-type superconductivity. The establishment of the MBE growth of $Bi_2Se_3$/monolayer $NbSe_2$ heterostructures and the demonstration of the crossover from 2D Ising- to 3D bulk Rashba-type superconductivity in these heterostructures will facilitate the investigations of the robust TSC phase in TI/Ising-type superconductor heterostructures.

The 2H phase of $NbSe_2$ bulk crystal is a well-studied *s*-wave superconductor[15]. Its unit cell consists of the Se-Nb-Se trilayer structure. The monolayer $NbSe_2$ has an out-of-plane mirror symmetry but breaks the in-plane inversion symmetry, which leads to Ising-type superconductivity in monolayer $NbSe_2$[7,8]. $Bi_2Se_3$ is a prototypical 3D TI, wherein a single Dirac cone exists at the Γ point of the first Brillouin zone [16,17]. The unit cell of $Bi_2Se_3$ consists of the Se-Bi-Se-Bi-Se QL structure. Since both $NbSe_2$ and $Bi_2Se_3$ are Se-based layered materials and share the same hexagonal lattice structure, it is natural to use MBE to synthesize both $NbSe_2$ and $Bi_2Se_3$ layers to form $Bi_2Se_3$/$NbSe_2$ heterostructures (Figs.1a and S1). However, as noted above, the superconductivity of MBE-grown $NbSe_2$ films usually fades away after the coverage of $Bi_2Se_3$ films. By carefully optimizing the growth protocols (see Methods), we successfully retained the superconductivity in $Bi_2Se_3$/monolayer $NbSe_2$ heterostructures. We first characterized $Bi_2Se_3$/monolayer $NbSe_2$ heterostructures using annular dark-field scanning transmission electron microscopy (ADF-STEM), as shown in Figs. 1b and S2. In addition to the prominent $Bi_2Se_3$ QL and $NbSe_2$ trilayer structures, we observed a BiSe bilayer with a cubic lattice structure at the $Bi_2Se_3$/$NbSe_2$ interface. This observation is consistent with prior studies on $Bi_2Se_3$/bulk $NbSe_2$ (Ref.[18]). The formation of this cubic BiSe layer is presumably stabilized by the large work function of the monolayer $NbSe_2$ underneath.

As noted above, monolayer $NbSe_2$ is an Ising-type superconductor[7,8], in which the broken in-plane inversion symmetry can generate an out-of-plane spin polarization under an effective Zeeman field (i.e. Ising-type spin-orbit interaction), which induces a Zeeman-type spin splitting at bulk valence band maxima located at K and K′ points with opposite signs (Fig.1c). Therefore, the inter-valley Cooper pairs in the monolayer NbSe$_2$ involve two electrons with locked opposite out-of-plane spins, which are referred to as Ising-type pairing. Superconductivity can also arise in materials with a Rashba-type band structure, where the spins of the electrons are pinned to the in-plane direction (Fig.1d). Cooper pairing between two such electrons with opposite in-plane spins is known as Rashba-type pairing. In our experiment, by depositing Bi$_2$Se$_3$ films on top of monolayer NbSe$_2$, Rashba-type conduction bands and Dirac SSs emerge and become superconducting due to the proximity effect. Therefore, Bi$_2$Se$_3$/monolayer NbSe$_2$ heterostructures provide a platform to study the crossover between Ising- and Rashba-type pairings in a single system.

We next performed *in-situ* ARPES measurements on $m$ QL Bi$_2$Se$_3$/monolayer NbSe$_2$ heterostructures with systematically increasing $m$. For monolayer NbSe$_2$ (i.e. $m = 0$), a hole pocket that crosses the Fermi energy near the Γ point was observed (Fig. S3), consistent with prior studies [19,20]. After deposition of 1 QL Bi$_2$Se$_3$ (i.e. $m = 1$) on monolayer NbSe$_2$, the ARPES band map shows a near-parabolic band dispersion, which is analogous to that of the 1 QL Bi$_2$Se$_3$ directly grown on epitaxial bilayer graphene substrates[13]. Compared to 1 QL Bi$_2$Se$_3$ grown on graphene, the 1 QL Bi$_2$Se$_3$/monolayer NbSe$_2$ is more electron-doped. The bulk conduction band minimum is at ($E$-$E_F$) ~ -0.55 eV (Figs. 2a and 2g). A gap of ~0.73 eV was observed for the $m = 1$ sample (Fig. S3). With increasing $m$, the hybridization gap between the top and bottom SSs decreases. The gap is ~0.18 eV for the $m = 2$ sample. Furthermore, we observed Rashba-type SSs for the $m = 2$ sample (Figs. 2b and 2h), implying symmetry-inequivalent top and bottom SSs (Ref. [13]). For the $m \geq 3$ samples, the SSs become gapless, a signature of crossover from 2D to 3D TI regimes (Figs. 2c and 2i)[21]. The crossover thickness in Bi$_2$Se$_3$/NbSe$_2$ heterostructures is found to be 3 QL. This is significantly lower than the 6 QL observed for $Bi_2Se_3$ films grown on other substrates including $NbSe_2$ bulk crystals [2,3,12-14]. With increasing $m$, the chemical potential gradually approaches the Dirac point (Figs. 2d to 2f and 2j to 2l).

In addition to the reduced 2D to 3D crossover thickness, the bulk conduction bands of $Bi_2Se_3$ films grown on monolayer $NbSe_2$ form quantum well (QW) states for $m \geq 2$ (Figs. 2b to 2f and 2h to 2l). Distinct from the QW states observed in $Bi_2Se_3$ films on sapphire[12], bilayer graphene[13], Si(111)[14], and $NbSe_2$ bulk crystals[2,3], the QW states in $m$ QL $Bi_2Se_3$/monolayer $NbSe_2$ heterostructures exhibit a Rashba-type band splitting, similar to a recent theoretical prediction based on $Bi_2Se_3$/monolayer $PtSe_2$ heterostructures[22]. To estimate the Rashba coupling strength $\alpha_R$ of the lowest QW states (labeled QW1 in Fig. 2), we defined the energy difference between the minimum of QW1 states and Rashba band crossing at the Γ point as $E_R$ and the momentum offset of two Rashba bands as $k_R$ (Fig. 2b). Consequently, the Rashba splitting parameter for a local parabolic dispersion is approximated by $\alpha_R = 2E_R/k_R$. For the $m = 2$ sample, $E_R \sim 0.019$ eV and $k_R \sim 0.038$ Å$^{-1}$, yielding $\alpha_R \sim 1.0$ eV·Å. The value of $\alpha_R$ in the 2 QL $Bi_2Se_3$/monolayer $NbSe_2$ heterostructure is comparable to that of the Rashba splitting of $Bi_2Se_3$ bulk conduction QW states[23-26]. With increasing $m$, $\alpha_R$ decreases monotonically and almost vanishes in the $m = 6$ sample, but the Rashba splitting of the second-lowest bulk QW states (labeled QW2 in Fig. 2) remains large in the $3 \leq m \leq 6$ samples (Figs. 2 and S4). The appearance of Rashba-type bulk QW bands and/or spin-nondegenerate Dirac SSs can together influence the occurrence of the superconducting proximity effect between $Bi_2Se_3$ and monolayer $NbSe_2$ films, as discussed below.

Next, we focused on transport measurements of $m$ QL $Bi_2Se_3$/monolayer $NbSe_2$ heterostructures. As shown in Fig. 2, the chemical potential crosses the Rashba-type bulk bands and/or spin-nondegenerate Dirac SSs, the electrons possess in-plane spin-polarization in both cases (Fig. 1d). When the superconducting proximity effect occurs between monolayer NbSe$_2$ film and $m$QL Bi$_2$Se$_3$ films, the Ising-type pairing in monolayer NbSe$_2$ film is suppressed and the Rashba-type pairing becomes progressively dominant with increasing $m$ in $m$ QL Bi$_2$Se$_3$/monolayer NbSe$_2$ heterostructures. Figure 3a shows the temperature dependence of the normalized longitudinal resistance $R$ of the $m$ QL Bi$_2$Se$_3$/NbSe$_2$ heterostructures with $0 \leq m \leq 6$. All these samples show superconductivity, with the superconducting transition temperatures ($T_c$) decreasing with increasing $m$. Here, we defined $T_c$ as the temperature at which $R$ drops to 50% of the normal state resistance found at $T = 5$ K. $T_c$ is found to decrease from ~2.8 K for the monolayer NbSe$_2$ to ~0.6 K for the $m = 6$ sample [27] (Fig. S5). Note that our ARPES band map of the $m = 1$ heterostructure shows a superposition of the electronic band structures of both 1 QL Bi$_2$Se$_3$ and monolayer NbSe$_2$ (Fig. S3). Since the valence band alignment of monolayer NbSe$_2$ does not change in the $m = 0$ and $m = 1$ samples, we can ignore the charge transfer effect between Bi$_2$Se$_3$ and monolayer NbSe$_2$ layers. Therefore, the observed superconductivity suppression (i.e. $T_c$ reduction) in our Bi$_2$Se$_3$/monolayer NbSe$_2$ heterostructures can be primarily attributed to the standard superconducting proximity effect [28] (Fig. S6) rather than the electron injection across the interface between Bi$_2$Se$_3$ and monolayer NbSe$_2$ layers [29].

We performed the magnetoresistance measurements on these Bi$_2$Se$_3$/monolayer NbSe$_2$ heterostructures under in-plane and out-of-plane magnetic fields (Figs. 3b and 3c). The upper critical magnetic fields $H_{c2,\parallel}$ and $H_{c2,\perp}$ are defined as the magnetic field at which $R$ is ~50% of the normal state resistance (Figs. S7 to S9). Figure 3b shows the reduced temperature ($T/T_c$) dependence of the normalized in-plane upper critical magnetic field ($H_{c2,\parallel}/H_P$) for the $m$QL Bi$_2$Se$_3$/monolayer NbSe$_2$ heterostructures with $0 \leq m \leq 6$. For the $m = 0$ sample, the ($H_{c2,\parallel}/H_P$)~($T/T_c$) curve near $T_c$ can be fitted by using $H_{c2,\parallel}/H_P \propto \sqrt{1 - T/T_c}$ for a 2D superconductor.

When 1 QL Bi$_2$Se$_3$ is deposited on top of monolayer NbSe$_2$, i.e. the $m = 1$ sample, the $(H_{c2,\parallel}/H_P) \sim (T/T_c)$ curve deviates slightly from this square-root-$T$ behavior, indicating a crossover from 2D to 3D superconductivity[30]. For the $m \geq 2$ samples, the $(H_{c2,\parallel}/H_P)\sim(T/T_c)$ curves can be fitted with $H_{c2,\parallel}/H_P \propto (1 - T/T_c)$ for a 3D superconductor. The linear temperature dependence of $H_{c2,\parallel}$ indicates that the orbital effect is dominant rather than the paramagnetic effect for the in-plane pair-breaking magnetic field in the Bi$_2$Se$_3$/NbSe$_2$ heterostructures with $m \geq 2$. Given the presence of the Rashba-type conduction bands and spin-nondegenerate Dirac SSs, the bulk Rashba-type superconductivity may appear in the $m \geq 2$ samples. Next, we extrapolated the value of $H_{c2,\parallel}/H_P$ at $T = 0$ K from the $(H_{c2,\parallel}/H_P) \sim (T/T_c)$ curves. In the monolayer NbSe$_2$, the extracted $H_{c2,\parallel}(T = 0K)$ is greater than 4.9 times the Pauli limit $H_P$, confirming the Ising-type pairing in monolayer NbSe$_2$. By depositing $m$ QL Bi$_2$Se$_3$ films on the monolayer NbSe$_2$, $H_{c2,\parallel}(T = 0K)/H_P$ is reduced to ~3.4 for $m = 1$ and ~2.0 for $2 \leq m \leq 6$. The value of $H_{c2}(T = 0K)/H_P$ greater than the factor of $\sqrt{2}$ in the $2 \leq m \leq 6$ samples provides further evidence that bulk Rashba-type superconductivity takes over from Ising-type behavior in Bi$_2$Se$_3$/monolayer NbSe$_2$ heterostructures[31].

The temperature dependence of $H_{c2,\perp}$ provides additional insights into the superconductivity in Bi$_2$Se$_3$/monolayer NbSe$_2$ heterostructures. We found that the $H_{c2,\perp} \sim (T/T_c)$ curves display a linear behavior in both monolayer NbSe$_2$ and $m$ QL Bi$_2$Se$_3$/monolayer NbSe$_2$ heterostructures due to the constraint of the orbital pair-breaking effect (Fig. 3c). As noted above, the Rashba-type superconductivity emerges in the $2 \leq m \leq 6$ samples, where the spins of two electrons are pinned to the in-plane direction (Fig. 1d). However, with an out-of-plane magnetic field, the orbital effect dominates, allowing the breaking of Cooper pairing. As a consequence, the enhancement of $H_{c2,\perp}$ was not observed in these bulk Rashba-type superconductors (Fig. 3c). We used the Ginzburg-Landau (GL) expression $H_{c2,\perp} = \Phi/(2\pi\xi_{GL,\perp}^2)$ to calculate the out-of-plane GL coherence length $\xi_{GL,\perp}$ from $H_{c2,\perp}(T = 0K)$. Here $\Phi$ is the magnetic flux quantum. The value determined for the monolayer NbSe$_2$, $\xi_{GL,\perp} \sim 14$ nm, is consistent with that found in bulk NbSe$_2$ crystals[4]. The value of $\xi_{GL,\perp}$ increases continuously with increasing $m$, suggesting the proximity-induced superconductivity occurs on both the interior and the top surface of the Bi$_2$Se$_3$ films even in the $m = 6$ sample. A proximity-induced superconducting gap on both the top and bottom Dirac SSs is a prerequisite for creating the TSC phase in TI/superconductor heterostructures[1]. For the $m = 6$ sample, $\xi_{GL,\perp}$ is found to be ~57 nm (Fig. 3c inset). A similar enhancement of $\xi_{GL,\perp}$ has been observed in proximity-induced superconductivity in Bi$_2$Te$_3$/bulk NbSe$_2$ heterostructures[4]. As a result, we posited that the proximity-induced superconducting gap occurs in both the Rashba-type bulk bands and spin-nondegenerate Dirac SSs of the Bi$_2$Se$_3$ layer. This leads to bulk Rashba-type superconductivity in the $m \geq 2$ samples. In addition to $\xi_{GL,\perp}$, the in-plane GL coherence length $\xi_{GL,\parallel}$ deduced from $H_{c2,\parallel}(T = 0K)$ also continuously increases with increasing $m$, further confirming the appearance of the proximity-induced superconductivity on the top surface of TI films (Fig. 3b inset).

To understand our transport results, we theoretically analyzed $H_{c2,\parallel}$ by considering the Zeeman coupling of monolayer NbSe$_2$. The Ising-type superconductivity in monolayer NbSe$_2$ serves as the Cooper pairing source for the bulk superconductivity in $m$QL Bi$_2$Se$_3$/NbSe$_2$ heterostructures. A strong Ising-type spin-orbit coupling ($\beta_{so} \sim 1.75$ meV) protects the superconducting states against the depairing term of the in-plane magnetic field, which gives rise to a large $H_{c2,\parallel}/H_P$. The enhanced paramagnetic limit[32] is described by

$$\text{Log}\left[\frac{T}{T_c}\right] = \text{Re}\left[\psi^{(0)}\left(\frac{1}{2}\right) - \psi^{(0)}\left(\frac{1}{2} + i\frac{\sqrt{\beta_{so}^2 + H_x^2}}{2\pi k_B T}\right)\right] \times \frac{H_x^2}{\beta_{so}^2 + H_x^2} \tag{1}$$

where $\psi^{(0)}(z)$ is the digamma function, $k_B$ is the Boltzmann constant, and $H_x$ is the Zeeman energy. We first used Eq. (1) to fit the $(H_{c2,\parallel}/H_P) \sim (T/T_c)$ curves of the monolayer NbSe$_2$ film with a single Ising-type spin-orbit coupling parameter $\beta_{so}$ and achieved the best fit with $\beta_{so} \sim 1.75$ meV (Fig. 4a). We next examined the influence of Rashba-type spin-orbit coupling on the upper critical field (see Eq. (8) in Methods), and with two fitting parameters $\beta_{so}$ and $\alpha_0 = \alpha_R k_F$. The Rashba parameter $\alpha_0$ reduces the upper critical field, thus, with a finite $\alpha_0$, one has to also increase $\beta_{so}$ to achieve a good fit. Figure 4b shows a fit for $\beta_{so} \sim 2.1$ meV with different $\alpha_0$, and the best fit is achieved with $\alpha_0 \sim 0.42 meV$, which is much smaller than $\beta_{so}$ i.e., $\alpha_0 < \beta_{so}/5$. Therefore, our theoretical calculations demonstrate that the 2D superconductivity in monolayer NbSe$_2$ carries Ising-type dominant Cooper pairs. For the $m = 1$ sample, the temperature dependence of the upper critical magnetic field deviates from either linear-$T$ (GL-3D limit) or square-root-$T$ (GL $-$ 2D limit) behavior, suggesting a complicated crossover behavior from a 2D Ising-type superconductor to a 3D bulk Rashba-type superconductor in this sample (Fig. 4c). The comparison between the theoretical model calculations and experimental data for the $m = 1$ sample is shown in Fig. 4c. The experimental data fall into the regime with the Rashba parameter $\alpha_0 \approx 0.7 \sim 1.1$ meV, which is comparable to $\beta_{so} \sim 1.75$ meV. The discrepancy between theory and experiment is attributed to the 3D orbital effects that are not captured in our 2D model. In the Supplementary Information, we discussed another mechanism of intervalley scattering that can also suppress $H_{c2,\parallel}$ due to the interfacial cubic BiSe bilayer (Fig.S10).

To identify the origin of the strong Rashba effect and the concurrent reduction of 2D to 3D TI crossover thickness in $m$QL $Bi_2Se_3$/BiSe/monolayer $NbSe_2$ heterostructures, we performed first-principles calculations to examine bulk QW bands and Dirac SSs. Capturing the entire $Bi_2Se_3$/BiSe/$NbSe_2$ heterostructure in first-principles calculations is challenging because a supercell commensurate with two different hexagonal lattices and a third cubic one is required. We instead examined two pairs of BiSe/$NbSe_2$ and $Bi_2Se_3$/BiSe. The electronic structure of BiSe/monolayer $NbSe_2$ shows that the band filling occupation of the monolayer $NbSe_2$ layer does not show any significant change (~$0.04e$ per $NbSe_2$ formula unit) (Fig. S11), consistent with our ARPES observation (Fig. S3). This insignificant charge transfer across BiSe/$NbSe_2$ leaves the hole carriers in the $NbSe_2$ layer uncompensated, which can still support the presence of superconductivity. Therefore, our following calculation of $Bi_2Se_3$/BiSe is a valid structural approximation of the $m$QL $Bi_2Se_3$/BiSe/monolayer $NbSe_2$ heterostructures.

Figure 4d shows the orbital-projected band structure of the 3QL $Bi_2Se_3$/BiSe heterostructure. For $-0.6$ eV $\leq (E - E_F) \leq -0.4$ eV, two Dirac SSs appear at different energies. The upper SS remains mostly intact, localized at the upper surface, while the lower SS drops to a lower energy level due to its interaction with the BiSe layer but has an increased decay length (Figs. 4e and S12). Therefore, we can attribute the energy separation of the two Dirac SSs to the interaction between the lower Dirac SS and the BiSe layer, which relieves the lower Dirac SS from further hybridizing with the upper Dirac SS. This explains why the critical thickness for the 2D to 3D TI crossover in $Bi_2Se_3$/monolayer $NbSe_2$ heterostructures is reduced. For $-0.25$ eV $\leq (E - E_F) \leq 0$ eV, three QW bands labeled as QW1, QW2, and QW3 appear, consistent with their *xy*-plane-averaged partial charge densities having 0, 1, and 2 nodes (Fig.4f). QW1 and QW2 are located at ~0.22 eV and ~0.39 eV above the Dirac point of the upper SS, respectively, in good agreement with

~0.23 eV and ~0.44 eV determined from the ARPES band map of the $m = 3$ sample (Figs. 2c and 2i). The calculated Rashba splitting $E_R$ of QW1 is ~0.009 eV (Fig. S13), which is also close to the value of $E_R$~0.011 eV measured by ARPES. Since $m$ QL $Bi_2Se_3$ films usually produce $(m-1)$ QW states to the conduction band minimum (Fig. S14), the three QW states suggest that the $m = 3$ sample is effectively equivalent to a freestanding 4QL $Bi_2Se_3$ sample (Fig. S15). In other words, the BiSe layer electronically resembles an additional $Bi_2Se_3$ layer. This is further confirmed by comparing the three QW states in the 3QL $Bi_2Se_3$/BiSe heterostructure with those in the freestanding 4QL $Bi_2Se_3$ sample (Fig. 4g). However, the 3QL $Bi_2Se_3$/BiSe heterostructure is different from the freestanding 4QL $Bi_2Se_3$ because the crystal potential of the former structure is asymmetric along $z$, as reflected from their partial charge densities plotted along $z$ in Fig. 4g. Therefore, the presence of the interfacial BiSe layer facilitates the formation of bulk Rashba-type QW bands with large $\alpha_R$ by introducing a compositional gradient along $z$ (Fig. S16).

In summary, we fabricated superconducting $Bi_2Se_3$/monolayer $NbSe_2$ heterostructures by MBE and found that an interfacial BiSe layer with a cubic lattice structure is formed at the interface between $Bi_2Se_3$ and monolayer $NbSe_2$ layers. By performing *in-situ* ARPES measurements, we observed the formation of gapless Dirac SSs in the 3 QL $Bi_2Se_3$/monolayer $NbSe_2$ heterostructure. Moreover, several bulk Rashba-type QW bands appear in $Bi_2Se_3$/monolayer $NbSe_2$ heterostructures with $Bi_2Se_3$ thickness greater than 2 QL. By performing magnetoresistance measurements, we found that both in-plane and out-of-plane upper critical magnetic fields of the superconducting $Bi_2Se_3$/monolayer $NbSe_2$ heterostructures are greatly reduced with increasing the $Bi_2Se_3$ film thickness, indicating the occurrence of a crossover from Ising- to bulk Rashba-type pairings. The successful synthesis of both an Ising-type superconductor and TI films by MBE and the observation of the crossover from Ising-type to bulk Rashba-type superconductivity may advance the exploration of the robust TSC phase in TI/Ising-type superconductor heterostructures.

**Methods**

**MBE growth of $Bi_2Se_3$/monolayer $NbSe_2$ heterostructures**

The $Bi_2Se_3$/monolayer $NbSe_2$ heterostructures used in this work were grown on the bilayer graphene terminated 6H-SiC (0001) substrates in a MBE chamber (ScientaOmicron) with a vacuum better than $2 \times 10^{-10}$ mbar. For the MBE growth of monolayer $NbSe_2$, the substrate temperature was kept at 500ºC. The monolayer $NbSe_2$ was post-annealed at 600 °C for 30 minutes to improve its crystallinity. Next, the $Bi_2Se_3$ films were grown on monolayer $NbSe_2$ at ~150ºC. This low growth temperature is crucial for achieving the superconductivity in $Bi_2Se_3$/monolayer $NbSe_2$ heterostructures. The MBE growth process was monitored by the reflection high-energy electron diffraction (RHEED) (Fig. S1). The samples were capped with a 10 nm Se layer to prevent their degradation during the *ex-situ* electrical transport measurements.

**ADF-STEM measurements**

The Aberration-corrected ADF-STEM measurements were performed on an FEI Titan³ G2 STEM operating at an accelerating voltage of 300 kV, with a probe convergence angle of 25 mrad, a probe current of 100 pA, and ADF detector angles of 42~244 mrad. See more STEM images of $Bi_2Se_3$/monolayer $NbSe_2$ heterostructures in Fig. S2.

**ARPES measurements**

The *in-situ* ARPES measurements were performed at room temperature after transferring the samples from the MBE chamber. A hemispherical Scienta DA30L analyzer was used, and the photoelectrons were excited by a helium discharged lamp with a photon energy of ~21.2 eV. The energy and angle resolution were set at ~10 meV and ~0.1°, respectively.

**Electrical transport measurements**

The electrical transport measurements were carried out in a Physical Property Measurement System (Quantum Design DynaCool, 2 K, 9 T), a Heliox $He^3$ refrigerator (Oxford Instruments, 400 mK, 6 T), and a dilution refrigerator (Bluefors, 20 mK, 9-1-1 T). Six-terminal mechanically defined Hall bars were used for electrical transport studies.

**First-principles calculations**

The first-principles calculations were performed using the Perdew-Burke-Ernzerhof (PBE) parametrization of the generalized gradient approximation exchange-correlation functional[33]. Interactions between electrons and nuclei were described by the projector augmented wave pseudopotentials[34] as implemented in the Vienna Simulation Package (VASP)[35]. Interlayer van der Waals interactions were accounted for using Grimme's DFT-D3 correction scheme[36]. Noncollinear spin textures and spin-orbit coupling were enabled following the native VASP implementation. Convergence parameters included the following: a plane-wave expansion energy cutoff of 500 eV, a residual force threshold of 0.01 eV/Å for relaxations (except for $m$ QL $Bi_2Se_3$ plus BiSe, see below), a 12×12×1 $k$-point grid for multilayer $Bi_2Se_3$ unit-cell calculations, and a 6×1×1 $k$-point grid for $Bi_2Se_3$ plus BiSe calculations.

For $Bi_2Se_3$ plus BiSe calculations, a 1×3$\sqrt{3}$ trilayer $Bi_2Se_3$ plus 1×5 cubic BiSe supercell was chosen, where $Bi_2Se_3$ adopted its native lattice constant of 4.14Å, requiring cubic BiSe to accommodate by sustaining a slight strain (1.6% and 2.2% in the lateral directions, considering BiSe's native lattice constant 4.21Å). This favoritism towards $Bi_2Se_3$ was to ensure the accuracy of the low-energy electronic structure of the heterostructure, dominated by $Bi_2Se_3$ bands (to be compared with experiments), and is also justified by the larger thickness of $Bi_2Se_3$ than BiSe. Structural relaxations for $Bi_2Se_3$ plus BiSe were prematurely terminated where residual forces were still ~0.2 eV/Å; at this stage, the softest degrees of freedom ($Bi_2Se_3$-BiSe interlayer separation) had already stabilized and resumed relaxation would strongly distort the BiSe layer through the formation of strong bonds between Bi (in BiSe) and Se (in $Bi_2Se_3$) wherever a Bi(BiSe)/Se($Bi_2Se_3$) pair happened to be close in the supercell. In a realistic structure including the underlying $NbSe_2$ and graphene substrate, such distortions would be suppressed since they would inflict additional energy penalty in interlayer adhesion and bending at the BiSe/$NbSe_2$ interface. Although a full relaxation with a near-commensurate supercell containing all structural components (i.e. $Bi_2Se_3$, BiSe, $NbSe_2$, and graphene) would be ideal, it would also require >1000 atoms (with spin-orbit coupling) and is therefore beyond our computational capabilities. For this reason, we used a flat $Bi_2Se_3$ plus BiSe structure in all electronic structure calculations.

**Theoretical calculations of critical temperature and upper critical magnetic fields**

To achieve the best fit for our experimental data of the monolayer $NbSe_2$ sample, we took an adjustment of $T_c$ with a 3% error from the experimental data into account. It means $T_c$ achieved in our experiments might be replaced by

$$T_c' = T_c + \Delta T_c \tag{2}$$

From the GL theory in the 2D limit, we knew that

$$b_0 \left(\frac{H}{H_P}\right)^2 = 1 - \frac{T(H)}{T_c'} = 1 - \frac{T(H)}{T_c+\Delta T_c} \tag{3}$$

It leads to

$$\frac{T(H)}{T_c+\Delta T_c} = 1 - b_0 \left(\frac{H}{H_P}\right)^2 \tag{4}$$

which can be rewritten as

$$\frac{T(H)}{T_c} = \frac{T_c + \Delta T_c}{T_c}\left[1 - b_0\left(\frac{H}{H_P}\right)^2\right] \quad (5)$$

for the fitting process. Here, there are two fitting parameters $b_0$ and $\Delta T_c$. During the fitting process, $|\Delta T_c|$ is less than $3\% T_c$.

In addition to the GL theory, we also adopted a model calculation to fit the in-plane upper critical field $H_{c2,\parallel}$ to reveal the role of the Ising-type spin-orbit coupling $\beta_{so}$ in our experiments. We considered the normal Hamiltonian of the monolayer NbSe$_2$ around K and K′ valleys [32] as

$$H_0 = \left(\frac{p^2}{2m_h} - \mu\right)\tau_0 s_0 + \beta_{so}\tau_z s_z + \alpha_R(p_x s_y - p_y s_x) + H_x s_x \quad (6)$$

where $m_h$ is the effective mass and $\tau, s$ are Pauli matrices in the valley and spin subspace, respectively. A phenomenological interaction Hamiltonian is considered to induce a spin-singlet inter-valley pairing state[32]. Note that the spin-triplet components induced by spin-orbit coupling are in order of $\beta_{so}/\mu \ll 1$, which can be ignored for simplicity[32].

To study the relationship between the in-plane upper critical magnetic field $H_{c2,\parallel}$ and the critical temperature $T/T_c$, we used the standard GL theory and solved the linearized gap equation[37]

$$\frac{1}{v_0} = -k_B T \sum_{i\omega_n, p} \text{Tr}[(\tau_x s_y) G_e(i\omega_n, p)(\tau_x s_y) G_h(i\omega_n, -p)] \quad (7)$$

where $v_0$ is the dimensionless attractive interaction strength, and $\omega_n = (2n+1)\pi k_B T$, $G_e(i\omega_n, p) = [i\omega_n - H_0(p)]^{-1}$ is the Matsubara Green's function for electrons, and $G_h = -G_e^*$ for holes.

We first considered a simple limit with $\alpha_R = 0$, then the solution of Eq. (7) gives rise to Eq. (1), which explains why the Ising-type superconductivity is resilient to an in-plane pair-breaking magnetic field. For the monolayer NbSe$_2$, the best fit is achieved with $\beta_{so} = 1.75$ meV. Under a weak magnetic field, the factor $H_x^2/(\beta_{so}^2 + H_x^2) \ll 1$ induces a large in-plane $H_{c2,\parallel} \gg H_P$.

Next, we included the Rashba-type spin-orbit coupling $\alpha_0$ which competes with $\beta_{so}$ (Ref.[32]), the solution of Eq. (7) becomes

$$\text{Log}\left[\frac{T}{T_c}\right] = \frac{1}{2}[C_0(\rho_-) + C_0(\rho_+)] + \frac{1}{2}[C_0(\rho_-) - C_0(\rho_+)] \times \frac{\beta_{so}^2 + \alpha_0^2 - H_x^2}{E_-E_+} \quad (8)$$

where $E_\pm = \sqrt{\beta_{so}^2 + (H_x \pm \alpha_0)^2}$ with $\alpha_0 = \alpha_R k_F$ and $k_F$ is the Fermi momentum, $\rho_\pm = \frac{1}{2}(E_+ \pm E_-)$. The kernel of the de-pairing function in Eq. (8) is given by $C_0(x) = \text{Re}\left[\psi^{(0)}\left(\frac{1}{2}\right) - \psi^{(0)}\left(\frac{1}{2} + i\frac{x}{2\pi k_B T}\right)\right]$ with $\psi^{(0)}(z)$ the digamma function. Note that Eq. (8) can be reduced to Eq. (1) in the main text by setting $\alpha_0 = 0$. The presence of $\alpha_R$ increases the depairing term $\rho_\pm$ which in turn decreases the in-plane $H_{c2,\|}$. However, through fitting the experimental data of the monolayer NbSe$_2$ sample, we found that the Rashba-type spin-orbit coupling energy $\alpha_0$ is much smaller than the Ising-type spin-orbit coupling $\beta_{so}$, i.e., $\alpha_0 < \beta_{so}/5$, and thus the Rashba effect can be ignored in the monolayer NbSe$_2$ (Fig. 4b). Finally, we used Eq. (8) to investigate the paramagnetic pair-breaking effect for the $m = 1$ sample. The moderate fit shows $\beta_{so} = 1.75$ meV and $\alpha_0 = 0.9$ meV (Fig. 4c), these two comparable spin-orbit couplings indicate the occurrence of the crossover from Ising-type (i.e. the $m = 0$ sample) to Rashba-type (i.e. the $m \geq 2$ sample) Copper pairs in Bi$_2$Se$_3$/monolayer NbSe$_2$ heterostructures.

**Acknowledgments**

We thank Yongtao Cui, K. T. Law, and Di Xiao for helpful discussions. This work is primarily supported by the Penn State MRSEC for Nanoscale Science (DMR-2011839). The electrical transport measurements and the sample characterization are partially supported by the NSF-CAREER award (DMR-1847811). The theoretical calculations and simulations are partially supported by the DOE grant (DE-SC0019064). Y. W. acknowledges the support from a startup grant from the University of North Texas. The MBE growth and the ARPES measurements were performed in the NSF-supported 2DCC MIP facility (DMR-2039351). The dilution refrigerator transport measurements at U. Washington are supported by the AFOSR award (FA9550-21-1-0177) and acknowledge the usage of the millikelvin optoelectronic quantum material laboratory supported by the M. J. Murdock Charitable Trust. D.R.H and N.A. acknowledge the support from the NSF-CAREER award (DMR-1654107). C. -Z. C. also acknowledges the support from the Gordon and Betty Moore Foundation's EPiQS Initiative (Grant GBMF9063 to C. -Z. C.).

**Author contributions:** C.-Z. C. conceived and designed the experiment. H. Y. performed the MBE growth, ARPES, and PPMS transport measurements with the help of Y.-F. Z, L-J. Z., R. Z, A. R. R., M. H. W. C., N. S., and C.-Z. C.. R. X. and H. Y. performed the Helium-3 transport measurements with the help of N. S. and C.-Z. C.. J. C. and X. X. performed the dilution transport measurements. Y. W. performed first-principles calculations. C. D. and J. A. R. prepared the bilayer graphene-terminated 6H-SiC(0001) substrates. D. R. H. and N. A. carried out the TEM measurements. L.-H. H. and C.-X. L. performed the theoretical simulations. H. Y., L.-H. H., Y. W., and C. -Z. C. analyzed the data and wrote the manuscript with inputs from all authors.

**Competing interests:** The authors declare no competing interests.

**Data availability:** The datasets generated during and/or analyzed during this study are available from the corresponding author upon reasonable request.

**Code availability:** The codes used in theoretical simulations and calculations are available from the corresponding author upon reasonable request.

**Figures and figure captions**

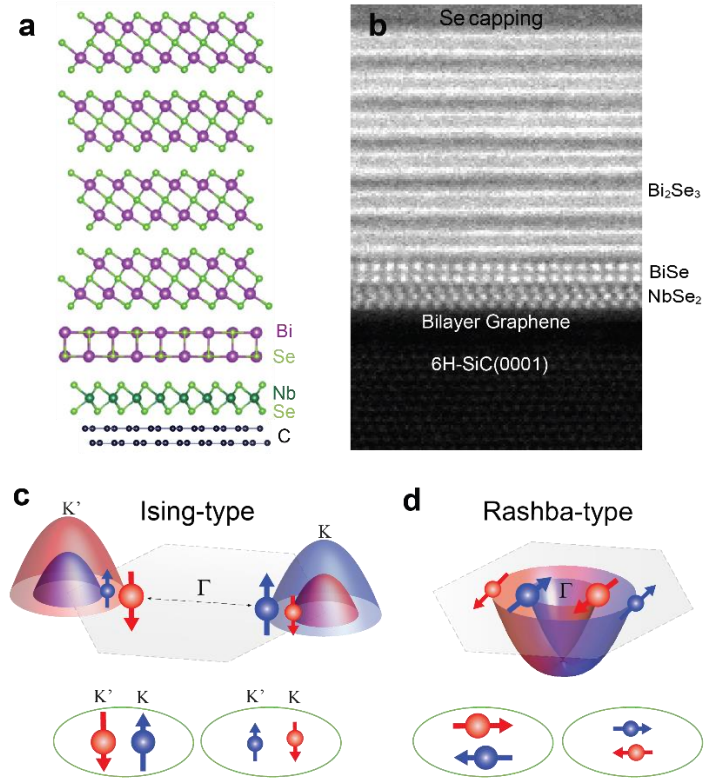

**Figure 1 | MBE-grown Bi$_2$Se$_3$/monolayer NbSe$_2$ heterostructures on epitaxial bilayer graphene.** (a) Schematic of the 4QL Bi$_2$Se$_3$/monolayer NbSe$_2$ heterostructure on bilayer graphene. (b) Cross-sectional ADF-STEM image of the 6 QL Bi$_2$Se$_3$/monolayer NbSe$_2$ heterostructure on bilayer graphene. The atomic structures of monolayer NbSe$_2$ and Bi$_2$Se$_3$ layers can not be simultaneously resolved, suggesting a misorientation between these two layers. (c, d) Schematic of the Ising-type (c) and Rashba-type (d) pairing symmetry in superconductors. For the Ising-type superconductor (e.g. monolayer NbSe$_2$), the electrons near the K and K′ valleys with spins pinned to the out-of-plane direction. However, for the Rashba-type superconductor (e.g. Bi$_2$Se$_3$/monolayer NbSe$_2$), the electrons near the Γ point with spins pinned to the in-plane direction.

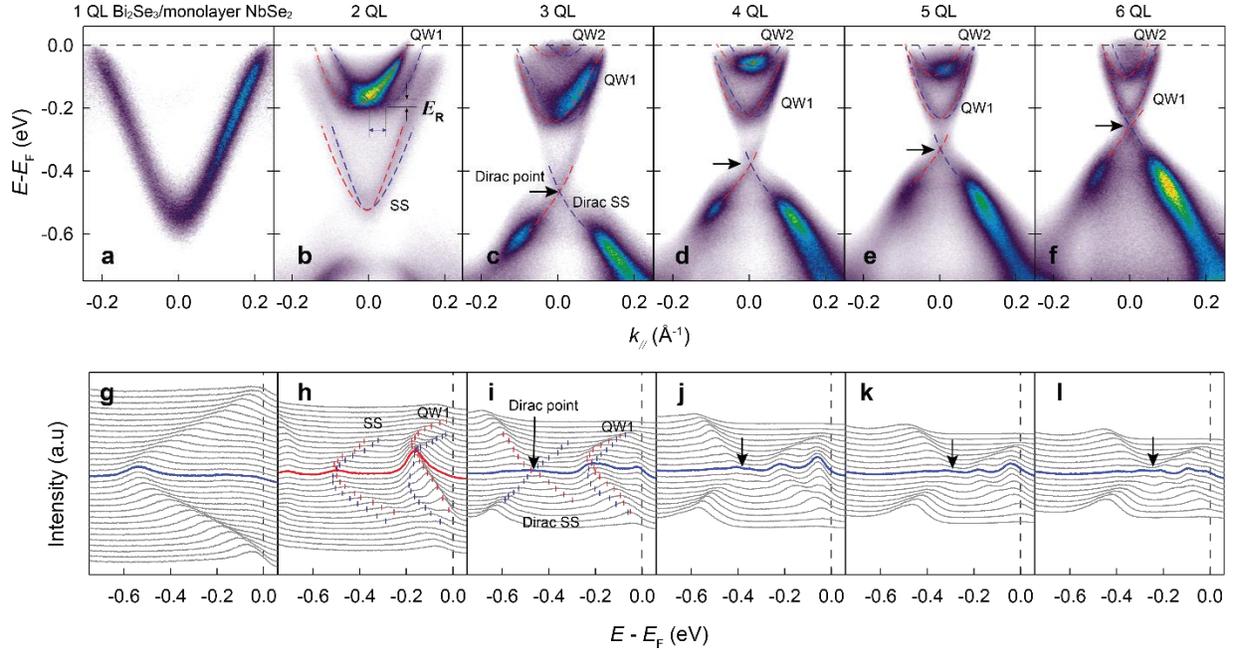

**Figure 2 | ARPES band maps of Bi$_2$Se$_3$/monolayer NbSe$_2$ heterostructures.** (a-f) ARPES results of $m$ QL Bi$_2$Se$_3$/monolayer NbSe$_2$ heterostructures with $m = 1$ (a), $m = 2$ (b), $m = 3$ (c), $m = 4$ (d), $m = 5$ (e), and $m = 6$ (f). (g-l) The corresponding energy distribution curves (EDCs) of the ARPES results shown in (a-e). The Rashba-type surface states (i.e. SS) and Rashba-type bulk QW states (i.e. QW1 and QW2) are highlighted by the red and blue dashed lines. The ARPES experiments were performed at room temperature. For the Bi$_2$Se$_3$ thickness greater than 3 QL, the Dirac point is formed and the second-lowest Rashba-type bulk QW states (i.e. QW2) appear.

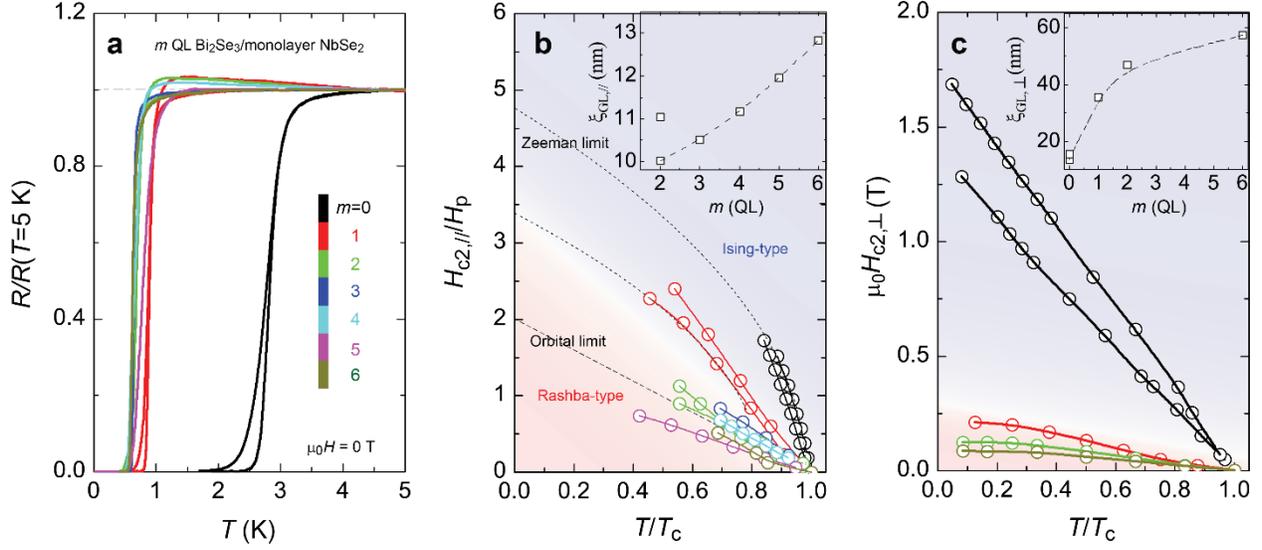

**Figure 3 | Crossover of Ising- to Rashba-type pairing symmetry in Bi$_2$Se$_3$/monolayer NbSe$_2$ heterostructures.** (a) $R-T$ curves of the $m$ QL Bi$_2$Se$_3$/monolayer NbSe$_2$ heterostructures under zero magnetic field. $R$ is normalized to the normal state resistance at $T=5$ K. The superconducting transition temperature $T_c$ is suppressed for $m \geq 1$. The enlarged $R$-$T$ curves are shown in Fig. S5. (b) The in-plane upper critical magnetic field normalized to the Pauli paramagnetic limit, $H_{c2,\parallel}/H_P$, as a function of the reduced temperature $T/T_c$ for the $m$ QL Bi$_2$Se$_3$/monolayer NbSe$_2$ heterostructures. (c) The out-of-plane upper critical magnetic field $H_{c2,\perp}$ as a function of the reduced temperature $T/T_c$ for the $m$ QL Bi$_2$Se$_3$/monolayer NbSe$_2$ heterostructures. Insets of (b and c) are the in-plane and out-of-plane Ginzburg-Landau coherence lengths $\xi_{GL,\parallel}$ and $\xi_{GL,\perp}$ as a function of $m$, respectively. Two $m$=0 samples are included in (a to c). Two $m$=1 and $m$=2 samples are included in (a to b).

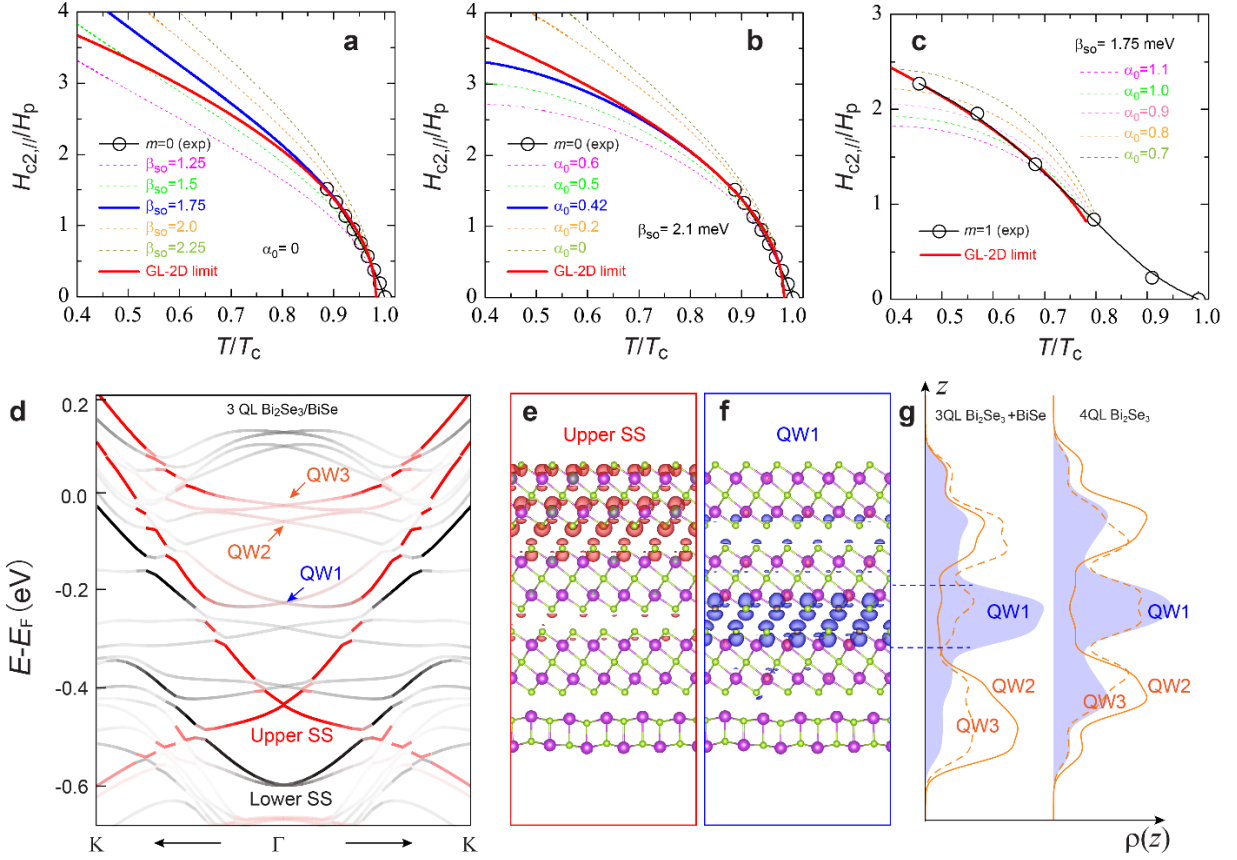

**Figure 4 | Theoretical calculations of $Bi_2Se_3$/monolayer $NbSe_2$ heterostructures.** (a to c) The calculated in-plane upper critical magnetic field normalized to the Pauli paramagnetic field $H_{c2,\parallel}/H_P$ as a function of the normalized temperature $T/T_c$ under different Ising-type spin-orbit coupling strength $\beta_{so}$ and Rashba-type spin-orbit coupling energy $\alpha_0$. In (a), for the $m = 0$ sample, we considered $\beta_{so}$ from 1.25 to 2.25 meV and $\alpha_0 = 0$. The best fit is approached with $\beta_{so} = 1.75$ meV and $\Delta T_c = -0.04$ K (blue solid line) for our experimental data (black circles), consistent with the GL-2D limit fit near $T_c$ (red solid line). In (b), we included a small $\alpha_0$ from 0 to 0.6 meV with fixed $\beta_{so} = 2.1$ meV for the $m = 0$ sample. The best fit is achieved for a weak $\alpha_0 = 0.42$ meV, indicating that the Ising-type Cooper pairs are still dominant in the $m = 0$ sample. In (c), the experimental data of the $m = 1$ sample can not be fitted by either linear-$T$ behavior (GL-3D limit) or square-root-$T$ behavior (GL-2D limit, red solid line). A moderate fit with two comparable spin-

orbit coupling parameters, i.e., $\alpha_0 \sim 0.9$ meV and $\beta_{so} \sim 1.75$ meV indicates the occurrence of the crossover from 2D Ising-type to 3D bulk Rashba-type superconductivity in the $m = 1$ sample. (d) The calculated band structure of 3QL Bi$_2$Se$_3$/BiSe heterostructure. The projections of the total wave function onto Bi orbitals on the upper and lower surfaces of 3QL Bi$_2$Se$_3$ are shown in red and black. More Rashba-type QW states in bulk conduction bands are marked by arrows. (e, f) Partial charge densities for the upper SS (e) and the first QW state (f). (g) Charge densities averaged in *xy*-plane plotted along *z*, for the three QW states in 3 QL Bi$_2$Se$_3$/BiSe heterostructure and the three QW states in a freestanding 4 QL Bi$_2$Se$_3$ film.